\documentclass[12pt,preprint]{aastex}
\shorttitle{Origin of X-shaped Radio Galaxies}
\shortauthors{Gopal-Krishna et al.}
\begin{document}

\title{The Origin of X-shaped Radio Galaxies: Clues from the
Z-symmetric Secondary Lobes}
\author{Gopal-Krishna\altaffilmark{1}}
\author{Peter L.\ Biermann}
\affil{Max-Planck Institut f{\"u}r Radioastronomie (MPIfR),
Bonn, Auf dem H{\"u}gel 69, 52121, Bonn, Germany}
\email{krishna@ncra.tifr.res.in; plbiermann@mpifr-bonn.mpg.de}
\and
\author{Paul J.\ Wiita}
\affil{Department of Physics \& Astronomy, MSC 8R0314,
Georgia State University,
Atlanta, GA 30303-3088}
\email{wiita@chara.gsu.edu}

\altaffiltext{1}{On leave from the National Centre for Radio Astrophysics (TIFR),
Pune University Campus, Pune - 411007, India}

\begin{abstract}

Existing radio images of a few X-shaped radio galaxies reveal  
Z-symmetric morphologies in their weaker secondary lobes 
which cannot be naturally explained by either the galactic merger or 
radio-lobe backflow
scenarios,  the two dominant
models for these X-shaped radio sources.  We show that
the merger picture can explain these morphologies provided one takes into
account that, prior to the coalescence of their supermassive black holes, 
the smaller galaxy releases
significant amounts of gas into the ISM of the dominant active galaxy. 
This rotating gas, whose angular momentum axis will typically not be aligned 
with the original jets, is likely to
provide sufficient ram pressure at a distance $\sim$10 kpc from the
nucleus to bend the extant jets emerging from the central engine, 
thus producing a Z-symmetry in the pair of radio lobes.
Once the two black holes have coalesced some
$10^7$ yr later, a rapid
reorientation of the jets along a direction close to that of
the orbital angular momentum of the swallowed galaxy relative
to the primary galaxy would create
the younger primary lobes of the X-shaped radio galaxy.  This picture naturally
explains why such sources typically have powers close to the FR I/II
break. We suggest that purely Z-symmetric radio sources are often
en route to coalescence and the concomitant emission of substantial 
gravitational
radiation, while X-shaped ones have already merged and radiated.

\end{abstract}


\section{Introduction}

Although the characteristic morphology of powerful (FR II) radio galaxies is a 
pair of radio lobes containing hotspots straddling the parent elliptical 
galaxy (Fanaroff \& Riley 1974), almost 10\% of the weaker FR II radio galaxies 
have a strikingly different morphology: they possess two misaligned pairs of
radio lobes of comparable extent (e.g.\ Leahy \& Parma 1992).
While the existence of such X-shaped, also called winged, radio galaxies (RGs)
was already recognized three decades ago (e.g.\ 3C 315, H{\"o}gbom \& Carlson 1974),
it is only very recently that they have shot into prominence. This is due 
to the growing realization that the origin of the X-shaped radio morphology 
may be a marker of the coalescence of two supermassive black holes (SMBH)
previously hosted by a pair of merging galaxies.  

Such an
interaction can lead to a rapid realignment of the angular momentum,
or spin-flip, of the active SMBH, accompanied by emission
of gravitational radiation of high intensity and angular momentum (e.g.\ 
Rottmann 2001; Zier \& Biermann 2001, 2002; Chirvasa 2002;
Merritt \& Ekers 2002; Biermann et al.\ 2003). 
In a related approach, Dennett-Thorpe et al.\ (2002) have concluded that a rapid 
jet reorientation could also arise from instabilities in the accretion disk,
though they agreed that minor mergers were likely to
be better explanations for the reorientations in 
3C 223.1 and 3C 403, sources they studied in detail. 
A somewhat related suggestion was that of Ekers et al.\ (1978),
who noted the possibility of a secular precession of
the jets due to a mismatch between their direction and the galaxy axis. 

An alternative and long discussed mechanism for the X-shaped morphology 
is based on the idea that the back-flowing plasma from the radio hotspots 
can be diverted due to pressure gradients in the external gas (e.g.\ Leahy 
\& Williams 1984; Worrall, Birkinshaw \& Cameron 1995). Arguing strongly 
for this general picture, Capetti et al.\ (2002) have recently pointed out
that X-shaped radio structures occur preferentially in galaxies of high
ellipticity and, furthermore, the secondary lobe pair (or wings) is 
usually closely aligned with the minor axis of the host galaxy, i.e., 
towards the steepest pressure gradient of the ambient medium as 
determined by the galaxy potential.  
Based on their two-dimensional hydrodynamical simulations
(which, however, assumed a reflection symmetry that artificially
enhances the numerically observed perpendicular outflow), 
Capetti et al.\ (2002) have further argued that 
the principal driving mechanism for the backflow diversion, appearing
as ordered outflows along the galaxy minor axes, is the over-pressure 
of the radio cocoon, which, aided by buoyancy forces, can 
conceivably lead to the secondary lobes growing as rapidly as the main lobe pair.

Clearly, a basic difference between the backflow-diversion and spin-flip 
scenarios is that whereas in the former case both the primary and the 
secondary lobe pairs are envisioned to form quasi-simultaneously, the creation
of the secondary (in the sense of lacking hotspots)
lobe pair in the spin-flip scenario occurs prior to the 
spin-flip event, while the currently active jets (i.e., after the spin-flip)
 feed the primary lobes,
which usually contain hotspots, and are thus classified as FR II sources.

\section{Rotating ISM and Z-symmetric jet bending}

We wish to draw attention to an apparently hitherto unnoticed morphological 
feature of  X-shaped RGs which is very difficult to reconcile with
either of the two proposed models mentioned above. Inspection of the
high resolution VLA maps of the prominent X-shaped RGs (XRGs) 3C 52 (Leahy \&
Williams 1984) and NGC 326 (e.g.\ Murgia et al.\ 2001)
shows that the ridge lines of the two extended secondary lobes are
not aligned with each other. The lobes are clearly offset from each other laterally
by roughly their widths (Fig.\ 1), with the inner edges of the secondary lobes 
aligned with each other and with the galactic nucleus. 
Thus, instead of extending in diametrically
opposite directions from the galactic nucleus, the secondary lobes exhibit
an approximately Z-shaped symmetry about the nucleus. 
Since a rather special aspect angle,
in addition to high angular resolution, is a prerequisite for being able to 
view such an offset, it is quite conceivable that a Z-symmetry of the
secondary lobe pair is actually a common feature of XRGs; another
possible example  is 3C 223.1 (Capetti et al.\ 2002). 

It is not obvious how the lateral offset of the secondary lobes, which
gives rise to their Z-symmetry, can arise naturally either in the spin-flip 
case, or in the backflow-diversion model, particularly when the axes of the
primary and secondary lobes are roughly orthogonal to each other, as they are 
in both of these cases.  In a variant of the 
original backflow-diversion picture (e.g.\ Leahy \& Williams 1984), the Z-symmetry 
of the secondary lobes could, 
conceivably, be enforced by the diversion of the backflow by a giant disk of 
denser material associated with 
the host galaxy and oriented roughly perpendicular to the primary radio axis
(Gopal-Krishna \& Wiita 2000). 
The existence of such {\it superdisks} has been inferred in several FR II 
radio galaxies from the sharp strip-like emission gaps observed between 
their radio lobes (Gopal-Krishna \& Wiita 2000). 
However,  superdisks also seem to provide only a partial
explanation for the Z-symmetries. Firstly, it is unclear how a nearly vertical 
infall of the back-flowing plasma onto the superdisk could lead to its 
diversion to just one side of the primary lobe axis. 
Furthermore,
this mechanism could not explain what makes the plume-like secondary lobes 
as long as, or even longer, than the directly powered main lobes.  
In at least some cases this cannot be due
to projection effects; e.g., for NGC 326 the jets are almost certainly
close to the plane of the sky (Murgia et al.\ 2001).  

To address these shortcomings, 
we propose a new model for the XRGs which encompasses an 
important modification to the spin-flip scenario. Basically, it takes into 
account the dynamical effect on the jets of the large-scale rotational field,
which is naturally set up in the ambient medium, as the captured galaxy, along 
with its SMBH, spirals in towards the core of the RG.  These gas motions
can bend the original jets (existing prior to the SMBH coalescence) into 
a Z-symmetric shape. Eventually, the
 coalescence of the two 
SMBHs produces a concomitant spin-flip of the active SMBH; its jets then 
create a new pair of (essentially unbent primary) radio lobes.


Z-symmetric bending of radio jets/lobes on kiloparsec (or larger) scales
is not a rare occurrence among radio galaxies, especially those of 
the FR I type (e.g.\ Miley 1980).
A likely explanation for such morphology can be given in terms of a rotating 
ISM which can bend sufficiently `soft' radio jets. 
A recent study of 21 nearby FR I RG hosts was able to detect significant
rotation in the H$\alpha$ emitting gas in the nuclear regions of 14 of them,
and most of the remainder were essentially face-on
(Noel-Storr et al.\ 2003); hence, significant amounts of rotating gas 
are very possibly ubiquitous in RG hosts.
The rapidly rotating ISM can be associated with the kpc-scale 
disks (or partial rings) which now have been found around the nuclei of many
early-type galaxies; in some cases these disks are able to survive long after 
the galaxy merger
as regular \ion{H}{1} structures (e.g.\ Oosterloo et al. 2002; 
Hibbard \& van Gorkom 1996).
Dense clouds containing both \ion{H}{1} gas and molecular gas at radial distances of 
$\sim$10 kpc from
the nucleus have  been clearly detected in the nearest radio galaxy, Cen A
(Schiminovich et al.\ 1994; Charmandaris, Combes \& van der Hulst 2000). 
The association of these dense clouds with some of the
stellar shells originally found by Malin (1979) and 
Malin, Quinn \& Graham (1983) shows that they 
are remnants of a merged gas-rich galaxy. 
Charmandaris et al.\ (2000) showed that
the association of these gaseous features with the stellar shells even at radial
distances of many kpc can be understood
within the standard picture of shell formation through galactic mergers
(e.g.\ Quinn 1984), if one takes 
into account that the ISM of the merged galaxy is clumpy, and hence not 
highly dissipative. 

The scenario of jet bending as a result of a jet--shell
interaction was originally proposed immediately 
after the discovery of optical shells (Gopal-Krishna \& Chitre 1983) and
discussed in 
detail in the context of Cen A (Gopal-Krishna \& Saripalli 1984).
This possibility was indirectly supported by numerical simulations 
of jets being bent by winds in both two-dimensional and three-dimensional
simulations (e.g.\  Loken et al. 1995), 
and is
strongly favored by the recent discoveries of \ion{H}{1} and 
H$_2$ components of some of the stellar shells in Cen A, as mentioned above. 
The measured 
radial velocities of the two diametrically opposite molecular clouds 
straddling the active nucleus
differ by 380 km s$^{-1}$  (Charmandaris et al.\ 2000). 
While not all of this is likely due to rotational motion around the
nucleus, it does indicate a 
rotational speed of $\sim$100 km s$^{-1}$ for the clouds at a radial 
distance of $\sim$15 kpc from the nucleus. 
The 
diameter of these clouds is $\sim$3 kpc, and thus their estimated masses
of $\sim 4\times 10^7 M_{\odot}$ imply an average gas density 
$\sim 4 \times 10^{-2} {\rm cm}^{-3}$.
The corresponding sideways dynamical pressure acting on the jets,
$p_{\rm dyn} = \rho_{\rm ISM} v^2_{\rm ISM}$,
is thus 
$\sim 6 \times 10^{-12}$ dyne, with $\rho_{\rm ISM}$ 
the mean mass density of the clouds and $v_{\rm ISM}$ their rotational velocity. 
Independent estimates of the energy densities in the inner radio lobes,
located roughly between 3 and 7 kpc from the galactic center,
range from
$1-9 \times 10^{-11}$ erg cm$^{-3}$ (Gopal-Krishna \& Saripalli 1984;
 Burns, Feigelson \& Schreier 1983; Feigelson et al.\ 1981).  In that
the energy density in the jets should have declined somewhat
by the time they propagate out to the distance of the clearly
detected moving shells, the ram pressure the latter could exert is 
sufficient to bend the jets enough to deflect them into the giant
lobes. 
Three-dimensional numerical simulations of light 
jets striking massive clouds with sizes several times the jet radius confirm
that stable bendings of $>$45$^{\circ}$ are possible
(Higgins, O'Brien \& Dunlop 1999; Wang, Wiita \& Hooda 2000).

For 
NGC 326, Murgia et al.\ (2001) determine energy densities
in the West wing of $\sim 8 \times 10^{-13}$ erg cm$^{-3}$ and in the
East wing of $\sim 3 \times 10^{-12}$ erg cm$^{-3}$ 
in the regions where
they are being bent.  
They point out that while the primary lobes are in rough equilibrium 
with respect to the surrounding intracluster medium,
which has a pressure of roughly $5 \times 10^{-12}$ dyne cm$^{-2}$
(Worrall \& Birkinshaw 2000),
the wings appear underpressured even with respect to the ICM. 
In the case of NGC 326 no direct evidence is available as yet for
the existence of gas clouds within the ISM (though its association 
with a dumbbell galaxy clearly points to a 
merger-prone environment, e.g.\ Colina \& de Juan 1995), 
but if they are present and have properties similar to those in 
Cen A, then they would easily be able to bend the secondary lobes. 
For the XRGs 3C 52 and 3C 223.1, evidence for a recent galaxy merger 
exists in the form of dust disks detected on the HST images 
(de Koff et al.\ 1996).

More generally, one can obtain rough constraints on the necessary parameters
of the rotating ISM with respect to those of the jets through the Euler 
equation applied to jet flows
(e.g.\ O'Donoghue, Eilek \& Owen 1993)
\begin{equation}
\rho v^2 / l_{\rm bend} \simeq \rho_{\rm ISM} v_{\rm ISM}^2 / l_{\rm pres},
\end{equation}
where $\rho$ is the density in the jet, $v$ the flow velocity through it,
$l_{\rm bend}$ the scale over which the jet bends, and
$l_{\rm pres}$ is the length scale over which the ISM applies the pressure
to the jet; this last quantity can be no less than the radius of the jet, $R_j \sim$1 kpc. 
Assuming 
the clouds in Cen A as reasonable
for the ISM and that the jet is sub-relativistic and made of ordinary proton-electron
plasma (pair plasma jets would be easier to bend),  we can 
normalize quantities as follows:
$l_{b,1} = l_{\rm bend}/10$ kpc, $l_{p,0} = l_{\rm pres}/ 1$ kpc,
$n_{\rm ISM,-1} = n_{\rm ISM}/0.1$ cm$^{-3}$, $v_{\rm ISM,2} = v_{\rm ISM}/100$ km s$^{-1}$.
We then have, 
$n v^2 \simeq 10^4 ~l_{b,1} ~l_{p,0}^{-1}~n_{\rm ISM,-1}~ v_{\rm ISM,2}^2$,
with the particle number density in the jet, $n$, 
in cm$^{-3}$ and $v$ 
in km s$^{-1}$.  If the jet velocity is semi-relativistic on
these scales, i.e., $v \sim 10^5$ km s$^{-1}$, we
require $n \sim 10^{-6}$ cm$^{-3}$ if bending is to be possible; 
this is a low, but certainly plausible, density. 
Somewhat higher densities in a bent jet can be accommodated if the
jet flow is slower or if the ISM ram pressure is higher; however, if the
jet thrust is very high, as is the case for powerful FR II sources,
even a clumpy ISM will have little impact on the jets (Higgins et al.\ 1999;
Wang et al.\ 2000). 

\section{Discussion and Conclusions}

As discussed above, 
$\sim 10^7$ years before the merger of the two SMBHs, a thick disk or
torus-like rotational field on the kpc scale can be expected to be 
established in the ISM of the RG, roughly along the orbital plane of the
captured galaxy's core. For the 
interesting case when the infalling black
hole with mass $M_2$ has a significant mass compared to the active SMBH
with mass $M_1$ (say $M_2 > 0.5 M_1$), 
the spin vector 
of the SMBH after the black holes' merger will be essentially along the 
orbital angular momentum vector of the captured black hole (Zier \& 
Biermann 2002); but even if $M_2 \simeq 0.05 M_1$, a significant
reorientation of the SMBHs spin axis to a direction predominantly along
the original orbital angular momentum is generally expected
(Merritt \& Ekers 2002).  Since, prior to the merger, the jets of the RG will
typically be oriented at a large angle to the 
orbital angular momentum vector, 
they would be subject to the sideways dynamic pressure
exerted by the rotating ISM 
and would, consequently, develop a Z-symmetric distortion unless their 
kinetic power is quite high.  
Such distorted radio structures are detected in several nearby galaxies with radio jets, 
such as M84 (Laing \& Bridle 1987),
Cen A (Gopal-Krishna \& Saripalli, 1984; 
Schiminovich et al.\ 1997)
and IRAS 04210+0400 (Holloway et al.\ 1996).

After the merger of the two SMBH, which is estimated to occur on a time scale
of $\sim 10^7$ yr on the pc scale (Zier \& Biermann 2001; Biermann et al.\ 2003), 
the two jets 
would be ejected from the merged SMBH along the new spin axis of the SMBH,  
and hence may face little bending from the ISM, whose rotation
axis may be derived from the orbit of the captured galaxy. The rotation
axes of the ISM and the merged black hole may be quite similar for
cases when the two black hole masses are not grossly unequal. At the
same time, the synchrotron plasma back-flowing from each hotspot would
expand into the nearest pre-existing, well established (low-density) secondary 
lobe. This injection of the back-flowing plasma into the
(Z-symmetric) secondary 
lobe pair would prolong their radio visibility, while the
pair of the newly created primary lobes continue to move
ahead.  In common with the standard spin-flip picture, one could 
see secondary lobes of greater length than the
primary lobes, something that is  quite implausible in any type
of pure back-flow model. 
Note that the diffusion of the back-flowing 
plasma from the primary lobes into the pre-existing secondary lobes would 
be facilitated through the buoyancy forces acting on the plasma.  This 
effect would be most pronounced 
in the case when the secondary lobes happen to extend roughly along the 
steepest ISM pressure gradient, i.e., along the minor axis of the active 
elliptical galaxy (e.g.\ Wiita 1978), thus explaining the correlation between secondary
lobe direction and optical galaxy morphology noted by Capetti et al.
(2002). 

Thus, to sum up, the scenario sketched here can provide a 
fairly natural explanation for the main morphological properties of X-shaped RGs 
(\S 1; Fig.\ 1). Simultaneously, it can account for the strong tendency of 
XRGs to be associated with relatively weak FR II radio sources 
(e.g.\ Leahy \& Parma 1992; Capetti et al.\ 2002). 
This is because, if the lobes created after the SMBH merger are of very low
radio luminosity, and thus have a pure FR I morphology, they would advance 
subsonically, like a plume of synchrotron plasma (e.g.\ Gopal-Krishna \& 
Wiita 1988, 2001; Bicknell 1995).  As soon as these advancing plumes  
approach the low-density channels associated with the pre-existing 
secondary lobes, they would be diverted into those lobes, instead of 
advancing straight ahead, as do the FR II lobes with jet-driven hot spots. 
Then no large-scale primary lobes would develop and one would only find a 
Z-symmetric FR I radio source.  On the other hand, very powerful FR II
sources will traverse the ISM quickly, and only a small amount
of cocoon plasma is likely to flow into the remnant lobes from
pre-realignment activity.  Thus any bright, reactivated regions of such lobes
will typically be much smaller than the
newly created  lobes; such sources are expected to have somewhat
broader than usual cocoons near the host galaxy but would not appear
as clearly X-shaped, or winged, RGs. 
Also recall that only a fraction of Z-symmetric XRGs will be oriented suitably that
we can observe the offsets of the secondary lobes, so the number of
XRGs with Z-symmetries is certainly significantly larger than the number identified so far.

It is also interesting to note that the primary radio lobes of 
both sources in Fig.\ 1 exhibit a mild ``C-symmetry''.  
Such  a jet bending can be understood in our model 
as arising from the linear momentum imparted to the 
radio-loud SMBH by the merged SMBH.   Also, just during the
merger of the two black holes, the accretion disk and the
base of the jet will be destroyed, but will be reestablished
a short time later; the consequences of the dip in the
radiation field and the realignment of the beamed emission
may entail observable effects.

In the model proposed here, the sources evolve along a
Z -- X morphological sequence. If basically correct, this picture can 
provide radio signatures of an impending merger of the SMBHs  
associated with two colliding galaxies.  A Z-symmetric 
radio morphology would often precede the SMBH merger, which would itself be manifested 
subsequently by the XRG phenomenon. Another important manifestation
predicted is the general relativistic emission of gravitational waves
during the SMBH merger (e.g.\ Rottmann 2001; Zier \& Biermann 2001; 
Chirvasa 2002;  Merritt
\& Ekers 2002; Biermann et al.\ 2003). We finally note that, in
addition to accounting for all the major observational attributes of
XRGs, our model can be potentially useful for inferring the sense of 
spin of the merged SMBH, since its spin would normally be dominated by the
orbital angular momentum vector of the infalling core of the captured galaxy
(e.g.\ Zier \& Biermann 2002) which, in turn, can be inferred from the 
observed bending of the secondary lobes into a Z-symmetric form. This
is an encouraging prospect, since the possibility of also inferring the spin 
direction of the SMBH from circular polarization measurements at centimeter
wavelengths has recently been discussed in the context of the cores of quasars
(Ensslin 2003).   

\acknowledgements
G-K thanks Prof.\ F.\ Combes for helpful discussions and 
acknowledges a senior Alexander von-Humboldt fellowship and hospitality 
at MPIfR.  
 PLB's work is mainly
supported through the AUGER theory
and membership grant 05 CU1ERA/3 through DESY/BMBF (Germany).
 PJW appreciates partial support from  RPE funds to PEGA at GSU and
hospitality at Princeton University. 
Thanks are also due to Herr W.\ Fussh{\"o}ller for artwork and to 
Drs.\ Leahy and Murgia for permission to reproduce their radio maps.


\newpage
Fig. 1.--- VLA maps of two prominent XRGs are shown: 3C 52 (Leahy \& Williams
1984, \copyright RAS) in the upper panel, 
and NGC 326 (Murgia et al.\ 2001, \copyright ESO)
in the lower panel, where several of the contours have
been omitted to improve clarity. The second image of 3C 52 on the right   
in the upper panel illustrates our model schematically. The
ridge lines of the two secondary lobes are marked with dotted lines. The
partial elliptical ring around the nucleus represents the postulated
shell-like features discussed in the text, whose direction of rotation 
is indicated by the arrow. 

\begin{figure}
\plotone{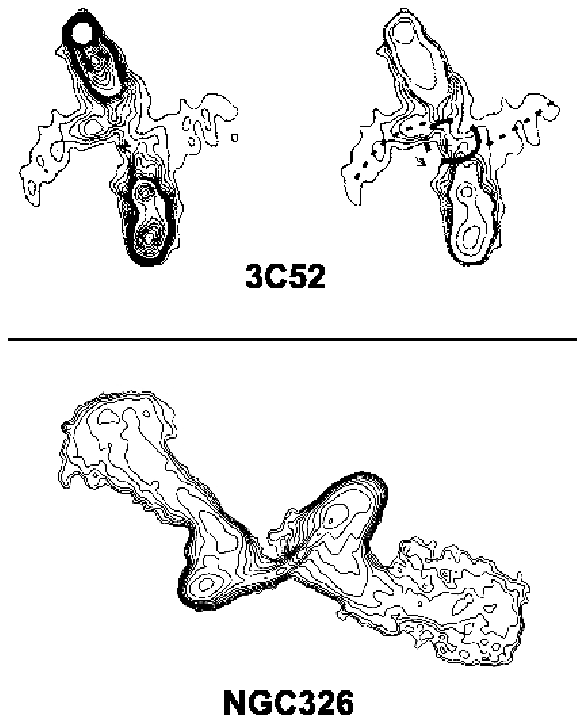}
\end{figure}

\end{document}